\documentclass[preprint,prb]{revtex4}
\usepackage{amsmath}
\usepackage{graphicx}
\usepackage{dcolumn}
\usepackage{bm}

\begin{document}

\begin{titlepage}

\title
{Impurity effects on the resonant Andreev reflection in a
finite-sized carbon nanotube system}

\author{Hui Pan, Tsung-Han Lin, and Dapeng Yu}
\affiliation{Department of Physics, State Key Laboratory for
Mesoscopic Physics, Peking University, Beijing 100871, P. R.
China}

\begin{abstract}
The influence of the impurity on the resonant Andreev reflection
through a normal-metal/carbon-nanotube/superconductor system is
studied theoretically. It is found that the resonant Andreev
reflection depends on the strength of the impurity and the length
of the armchair nanotube. The impurity which breaks the
electron-hole symmetry of the nanotubes greatly reduces the
resonant Andreev reflection. The symmetry broken depends
distinctly on the impurity strength. The impurity effects on the
Andreev reflection current at different bias are also studied.

\noindent {PACS number(s)}: 72.80.Rj, 73.23.Ad, 73.63.Fg
\end{abstract}
\maketitle
\end{titlepage}

Carbon nanotubes have been the subject of an increasing number of
experimental and theoretical studies due to their
quasi-one-dimensional structure and unique electronic
property.\cite{Saito1} The perfect carbon nanotube is predicted to
be either metallic or semiconducting sensitively depending on its
diameter and chirality, which is uniquely determined by the chiral
vector $(n,m)$, where $n$ and $m$ are
integers.\cite{Saito1,Mintmire,Hamada,Saito2} Experimental and
theoretical studies have indicated that the electronic and
transport properties of carbon nanotubes can be substantially
modified by point defects such as the substitutional
impurities.\cite{Bochrath,Chico1,Chico2,Chico3,Anantram,Kostyrko,Choi}
One of recent interests concentrates on the electron transport
through hybrid nanotube system. Experiments about some hybrid
systems including nanotube-based magnetic tunnel
junctions\cite{Tsukagoshi} and superconducting
junctions\cite{Kasumov,Morpurgo} have been successfully
fabricated. The electrical transport transport about the carbon
nanotube quantum dot in the Kondo regime coupled to a normal and
a superconductor has also been reported.\cite{Graber} The
theoretical investigation of transport properties of these hybrid
nanotube devices is of great importance, not only for their basic
scientific interest, but also aiming at the design of novel
nanodevices.

The resonant Andreev reflections in the
superconductor/carbon-nanotube devices has been theoretically
studied.\cite{Wei} The proximity effect in
superconductor/carbon-nanotube/superconductor (S/CNT/S) tunnel
junctions has also been studied theoretically.\cite{Jiang}
However, the effects of the impurity, such as the substitutional
boron (nitrogen) and the vacancy, in these systems are not
considered. In this paper, the effects of the substitutional boron
(nitrogen) and the vacancy on the resonant Andreev reflection in
the hybrid normal-metal/carbon-nanotube/superconductor (N/CNT/S)
system are theoretically studied. In such a system, the specific
molecular orbital plays an important role. By combing standard
nonequilibrium Green's function (NGF)
techniques\cite{Yeyati1,Yeyati2,Sun} with a tight-binding
model,\cite{Nardelli1,Nardelli2} we have analyzed quantum
transport properties of the N/CNT/S system with the
substitutional boron (nitrogen) and the vacancy. The Andreev
reflection through the finite-sized carbon nanotube depends on
the impurity strength and the tube length. The substitutional
impurity, such as the boron and nitrogen, reduces the resonant
Andreev reflection greatly. However, the impurity with very large
strength, such as the vacancy, does not reduce the resonant
Andreev reflection. The dependence of the Andreev reflection
current with the gate voltage is also studied.

We assume that the system N/CNT/S under consideration is described
by the following Hamiltonian:
\begin{equation}
H=H_{L}+H_{R}+H_{tube}+H_{T},
\end{equation}
where
\begin{equation}
H_{L}=\sum_{k,\sigma}(\epsilon_{L,k}^{0}-ev_{L})a_{L,k\sigma}^{\dag}a_{L,k\sigma},
\end{equation}
\begin{equation*}
H_{R}=\sum_{p,\sigma}\epsilon_{R,p}^{0}a_{R,p\sigma}^{\dag}a_{R,p\sigma}
+\sum_{p}[\Delta^{*}a_{R,p\downarrow}^{\dag}a_{R,-p\uparrow}
+\Delta a_{R,-p\uparrow}^{\dag}a_{R,p\downarrow}],
\end{equation*}
\begin{equation*}
H_{tube}=\sum_{\langle
i,j\rangle,\sigma}[\gamma_{0}c_{i\sigma}^{\dag}c_{j\sigma}
+H.c.]-\sum_{i,\sigma}ev_{g}c_{i\sigma}^{\dag}c_{i\sigma}+Uc_{0}^{\dag}c_{0},
\end{equation*}
\begin{equation*}
H_{T}=\sum_{k,\sigma,i}[t_{L}a_{L,k\sigma}^{\dag}c_{i\sigma}+H.c.]
+\sum_{p,\sigma,i}[t_{R}e^{iev_{R}\tau}a_{R,p\sigma}^{\dag}c_{i\sigma}+H.c.],
\end{equation*}
where $H_{L}$ describes the noninteracting electrons in the left
normal-metal lead, $a_{L,k\sigma}^{\dag}(a_{L,k\sigma})$ are the
creation (annihilation) operators of the electron in the left
lead, and $v_{L}$ is the voltage of the left lead. $H_{R}$
describe the right superconducting lead with the energy gap
$\Delta$. The nanotube Hamiltonian $H_{tube}$ is described by the
tight-binding model with one $\pi$-electron per atom at $i$ site.
The sum in $i$, $j$ is restricted to nearest-neighbor atoms, and
the bond potential $\gamma_{0}=-2.75eV$, which is used as the
energy unit. This model is known to give a reasonable,
qualitative description of the electronic and transport
properties of carbon nanotubes.\cite{Nardelli1,Nardelli2} For
simplicity, the intra-tube electron-electron Coulomb interaction
has been neglected. $v_{g}$ is the gate voltage which controls
the energy levels in the CNT. The pointlike defect is defined by
setting site energy equal to $U$ at one of the sites of the unit
cell, and various strengths represent typical substitutional
impurities or vacancy.\cite{Kostyrko}. According to former
tight-binding and \textit{ab inito}
calculations,\cite{Kostyrko,Choi} we set the strength $U=3$,
$-5$, and $10000$ to simulate the substitutional boron, nitrogen
and vacancy, respectively. $H_{T}$ denotes the tunneling part of
the Hamiltonian, and $t_{L,R}$ are the hopping matrix. It is
convenient to introduce the $2\times 2$ Nambu representation in
which the Green's function can be expressed by
\begin{equation}
G^{r,a}(\tau,\tau')=\mp i\theta(\tau\mp\tau')
\sum_{ij}\left(\begin{array}{cc}
\langle{c_{i\uparrow}(\tau),c_{j\uparrow}^{\dag}(\tau')}\rangle
&\langle{c_{i\uparrow}(\tau),c_{j\downarrow}(\tau')}\rangle \\
\langle{c_{i\downarrow}^{\dag}(\tau),c_{j\uparrow}^{\dag}(\tau')}\rangle
&\langle{c_{i\downarrow}^{\dag}(\tau),c_{j\downarrow}(\tau')}\rangle
\end{array}\right),
\end{equation}
The retarded Green's function of the nanotube is calculated
directly using a tight-bingding model via
\begin{equation}
g^{r}(\epsilon)=\left(\begin{array}{cc}
\frac{1}{\epsilon-H_{tube}+i0^{+}} & 0\\
0 & \frac{1}{\epsilon+H_{tube}+i0^{+}}
\end{array}\right).
\end{equation}
Using the standard NGF technique\cite{Yeyati1,Yeyati2,Sun}, the
Green's functions are obtained as
\begin{equation}
G_{11}^{r}(\epsilon)=[(g_{11}^{r}(\epsilon))^{-1}
 +\frac{i}{2}\Gamma_{L}+\frac{i|\epsilon|}{2\sqrt{\epsilon^{2}-\Delta^{2}}}\Gamma_{R}
 +\frac{\Delta^{2}}{4(\epsilon^{2}-\Delta^{2})}A^{r}(\epsilon)]^{-1},
\end{equation}
\begin{equation*}
G_{12}^{r}(\epsilon)=G_{11}^{r}(\epsilon)
 \frac{\Delta}{2\sqrt{\epsilon^{2}-\Delta^{2}}}\Gamma_{R}A^{r}(\epsilon),
\end{equation*}
\begin{equation*}
A^{r}(\epsilon)=[(g_{22}^{r}(\epsilon))^{-1}
 +\frac{i}{2}\Gamma_{L}+\frac{i|\epsilon|}{2\sqrt{\epsilon^{2}-\Delta^{2}}}\Gamma_{R}]^{-1}.
\end{equation*}
where $\Gamma_{L,R}$ are the appropriate linewidth functions
describing the coupling of the CNT to the respective leads. Here
$G_{11}$ and $G_{12}$ are the retarded Green's functions of the
CNT, which include the proper self-energy of the
leads.\cite{Yeyati1,Yeyati2,Sun} Then the current and probability
of the Andreev reflection are given by
\begin{equation}
I_{A}=\frac{2e}{h}\int d\epsilon
[f_{L}(\epsilon+ev_{L})-f_{L}(\epsilon-ev_{L})]T_{A}(\epsilon),
\end{equation}
\begin{equation}
T_{A}(\epsilon)=Tr[\Gamma_{L}G_{12}^{r}(\epsilon)\Gamma_{L}G_{12}^{r\dag}(\epsilon)],
\end{equation}
where $f_{L,R}$ denote the Fermi functions of the left and right
leads, respectively. Clearly, the conventional tunneling is
completely forbidden for $V<\Delta$, and only the Andreev
reflection exists. In the following numerical calculations, we
discuss in detail the Andreev reflection at zero temperature in
the case of $V<\Delta$. We set (1) the temperature
$\mathcal{T}=0$, (2) the voltage of the right lead $v_{R}=0$ due
to the gauge invariance, and carry out all calculations with
$\Delta=1$, and $\Gamma_{L}=\Gamma_{R}=0.02$ in units of
$\hbar=e=1$.

In the following, the probability $T_{A}$ and current $I_{A}$ of
Andreev reflection for the N/CNT/S hybrid system are calculated.
In Fig. 1, $T_{A}$ is plotted as a function of the incident
electron energy for nanotubes with different lengths. For the
comparison, the conductance $G$ of the N/CNT/N system with $L=3$
is plotted in Fig. 1(a). The conductance peaks reflect the band
structures of the finite-sized nanotubes, because the resonant
states are close to the eigenvalues for small coupling $\Gamma$.
In the $\pi$-electron tight-binding model, the defect-free
nanotubes have complete electron-hole symmetry with their Fermi
levels at zero.\cite{Chico2} Then the resonant states are
symmetric around the Fermi energy $E_{F}=0$. The Andreev
reflection probability for the N/CNT/S system with $L=3$ are
plotted in Fig. 1(b). The resonant Andreev reflection reflects
the distribution of the resonant states of the nanotube systems.
Although the resonant peaks are similar to those in a normal
N/CNT/N system, they are different from the conventional resonant
tunneling, because the conventional tunneling is completely
forbidden for $|E|<\Delta$. In fact, these peaks come from the
Andreev reflections. For the finite-sized pristine nanotubes, the
resonant states are symmetric around the Fermi energy as
mentioned above. Furthermore, the chemical potential of the right
superconducting lead $\mu_{R}=0$ is lined up with the Fermi
energy of the nanotube. It is just located in the middle of two
symmetric states with energy $\epsilon_{i}$ and $-\epsilon_{i}$.
When the electron incident from the left lead has the energy
$\epsilon_{i}$ corresponding the $i$th state of the nanotube, a
hole can propagate back to the state with the energy
$-\epsilon_{i}$. Then a Cooper pair creates in the right
superconducting lead because of Andreev reflection. The
conductance for the nanotube with $L=4$ is plotted in Fig. 1(c).
It is evident that both positions and heights of the resonant
peaks depend on the nanotube length. There is one resonant peak
with the amplitude of two units at the Fermi energy, which is
absent for the nanotube with $L=3$. Similarly, there is a
resonant Andreev peak at the Fermi energy for the N/CNT/S system
with $L=4$ as shown in Fig. 1(d). This is attributed to the
electronic properties of the nanotubes. The band structure of
armchair nanotubes consists of two non-degenerate bands that cross
the Fermi level at $k_{F}=2\pi/3a$, with lattice constant $a$.
Finite size effects in carbon nanotubes lead to the quantization
of the energy levels. In general, one resonant peak appears at
the Fermi energy with $L=3N+1$ ($N$ denoting the number of carbon
repeat units), because $k_{F}=2\pi/3a$ is now an allowed wave
vector, a large conductance exists due to a crossing of two
resonant states at the Fermi energy.\cite{Orlikowski} For other
lengths, $k_{F}$ is not an allowed wave vector and no resonant
state exists at the Fermi level, thus conductance is much smaller
due to the energy gap between the resonant states. These are
referred to as on-resonance and off-resonance of the Andreev
reflection, respectively.

The impurity, such as the substitutional boron (nitrogen) and the
vacancy, can greatly change the electronic structure of the
nanotubes and then the transport properties. In general, the
impurity increases the normal reflection.\cite{Kostyrko} The
substitutional boron or nitrogen impurity in the infinite carbon
nanotube lead to a quasibound state near the lower or upper
subbands.\cite{Choi} However, it is quite different for
finite-sized carbon nanotubes. New resonant state appears when
the incoming electron energies match those of the quasibound
states induced by the impurity. The substitutional boron effects
on the conductance of the N/CNT/N system with $L=3$ are clearly
shown in Fig. 2(a). Compared with Fig. 1(a), the impurity leads
to one new peak below the Fermi energy. The resonant states for
the conduction and valence bands are quite different due to the
impurity, because the electron-hole symmetry, present in perfect
nanotubes within the $\pi$-band approximation, is
broken.\cite{Chico3} The positions of the resonant peaks is not
symmetric around the Fermi energy $E_{F}=0$ due to the symmetry
broken. The corresponding Andreev reflection probability is
plotted in Fig. 2(b). It can be clearly seen the substitutional
boron greatly reduces the Andreev reflection probability. Because
of the broken of the electron-hole symmetry mentioned above, the
energy levels of the finite-sized nanotubes are not symmetric
around $E_{F}=0$. Because the energy level $\epsilon_{i}$ does
not have a corresponding one with $-\epsilon_{i}$, the condition
for the Andreev reflection has been broken. When the electron
incident from the left lead has the energy $\epsilon_{i}$
corresponding to the $i$th state, a hole can not propagate back
to the state with the energy that is different from
$-\epsilon_{i}$. The Andreev reflection is then reduced
distinctly by the impurity. Fig. 2(c) shows the conductance for
the nanotube with $L=4$. It is seen that the original peak at the
Fermi energy with the height of two units are splitted into two
ones with the height of one unit each. One of the two peaks is
still at the Fermi energy and the other one is above the Fermi
level. The reason is that the single impurity breaks the mirror
symmetry planes containing the tube axis, and then the two
resonant states at the Fermi energy are now splitted into two
ones. Fig. 2(d) shows the corresponding Andreev reflection
probability for $L=4$. The Andreev reflection for $L=4$ is
stronger than that for $L=3$, because the resonant states are
more symmetric for $L=4$. Furthermore, there is also one peak at
the Fermi energy due to the reasons mentioned above.

The conductance and the corresponding Andreev reflection
probability for the nanotube with the substitutional nitrogen are
plotted in Fig. 3. A substitutional nitrogen has similar effects
on the conductance of the N/CNT/N system as the boron, but it
induces one resonant state above the Fermi energy. The resonant
state associated with boron or nitrogen is analogous to the
acceptor or donor state in semiconductors.\cite{Choi} The
nitrogen impurity also reduces the Andreev reflection, because it
breaks the symmetry of the resonant states of the finite-sized
nanotubes. Because the electron-hole spectra becomes more
symmetrical with stronger $U$, the degree of the symmetry broken
is slighter. The reduction is not so heavily as that caused by
the boron. Then the Andreev reflection becomes larger again for
stronger $U$. Fig. 4 shows the effects of the vacancy on the
conductance and the Andreev reflection. For the very large $U$,
as shown in Fig. 4(a), the resonant states become symmetric
again. However, it is quite different from Fig. 1(a). One new
peak emerges near the Fermi energy, which is quite different from
that of the perfect nanotubes. The position of the resonant state
caused by the impurity approaches to the Fermi energy for a very
strong $U$.\cite{Kostyrko} Then, it is expected that the vacancy
dose not reduce the Andreev reflection, which can be seen from
Fig. 4(b).  For the off-resonant nanotube, the vacancy causes the
appearance of a resonance state at the Fermi energy, where it is
originally zero for the perfect nanotube. For the on-resonant
nanotube, the vacancy induces a new resonant peak at the Fermi
energy as shown in Fig. 4(c). The original one at the Fermi
energy is splitted into two smaller ones around the Fermi energy.
Because the electron-hole symmetry is recovered, the Andreev
reflection becomes stronger again as shown in Fig. 4(d).

The Andreev reflection current versus the gate voltage is also
investigated at different bias. As shown in Fig. 5(a)-(c), the
current for the nanotube with $L=3$ exhibits a single series of
peaks with different spacings at very small bias, which reflects
the band structure of the finite-sized nanotubes with the
substitutional impurity. At a larger bias, as shown in Fig.
5(d)-(e), a series of extra peaks emerges. And they are located
in the middle of the original ones at the low bias. Furthermore,
the amplitude of some original peaks become larger and reaches a
height three times its value at the low bias. The reason is that
more energy levels contribute to the Andreev reflection at high
bias. When $V>\Delta\epsilon/2$, a series of extra peaks can
emerge between the original ones. And when $V>\Delta\epsilon$,
the height of the original peaks can become about three times the
original value.\cite{Sun} Due to the different energy spacings,
for some high bias, both of the conditions mentioned above can be
satisfied. The impurity effects on the Andreev reflection current
for the tube with $L=4$ are clearly shown in Fig. 6. There are
more peaks in the current at the same bias compared with Fig. 5.
For nanotube with longer length, the spacing between energy
levels become smaller. Because more resonant states contribute to
the current at the same bias, there are more peaks in the Andreev
reflection current. The complicated resonance patterns with
different kinds of peaks depend on the bias, the level spacing of
the CNT. It is also found that the exact location of the impurity
has less influence on this significant feature than the strength
of the impurity.

In summary, the probability and the current of the Andreev
reflection for the N/CNT/S hybrid system are studied in detail.
The Andreev reflection exhibits the on-resonance and
off-resonance behavior at the Fermi energy for the nanotubes with
$L=3N+1$ and other lengths, respectively. The substitutional
boron, nitrogen and vacancy are simulated by different impurity
strength $U$. The substitutional boron or nitrogen breaks the
electron-hole symmetry of the nanotube and reduces the Andreev
reflection greatly. The symmetry broken and the reduction depend
on the impurity strength distinctly. The vacancy with very strong
impurity strength does not reduce Andreev reflection, because it
keeps the electron-hole symmetry. It can change the off-resonance
to the on-resonance of the Andreev reflection at the Fermi
energy. The Andreev reflection current shows a complicated
behaviour at different bias, which reflect the complex band
structure of the finite-sized nanotubes with the impurity.

Financial support from National Natural Science Foundation (Grant
No. 90103027 and 50025206) and the National "973" Projects
Foundation of China are gratefully acknowledged.



\begin{center}
{\bf Figure Captions}
\end{center}

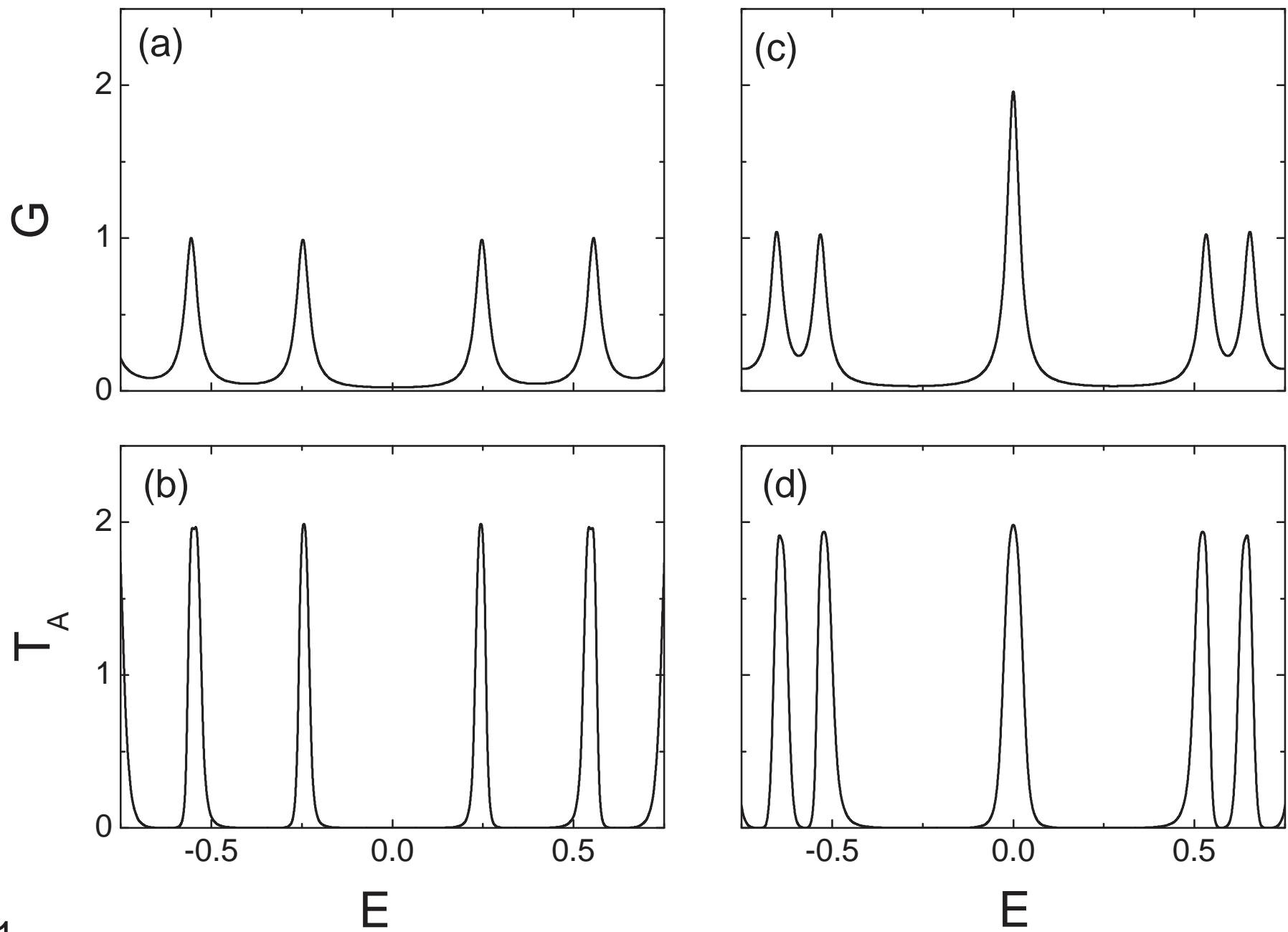
Fig. 1. (a) and (b) are the conductance $G$ and Andreev
reflection probability $T_{A}$ as a function of electron energy
$E$ for the N/CNT/N and N/CNT/S systems of $L=3$, respectively.
(c) and (d) are the corresponding $G$ and $T_{A}$ for the systems
of $L=4$, respectively.

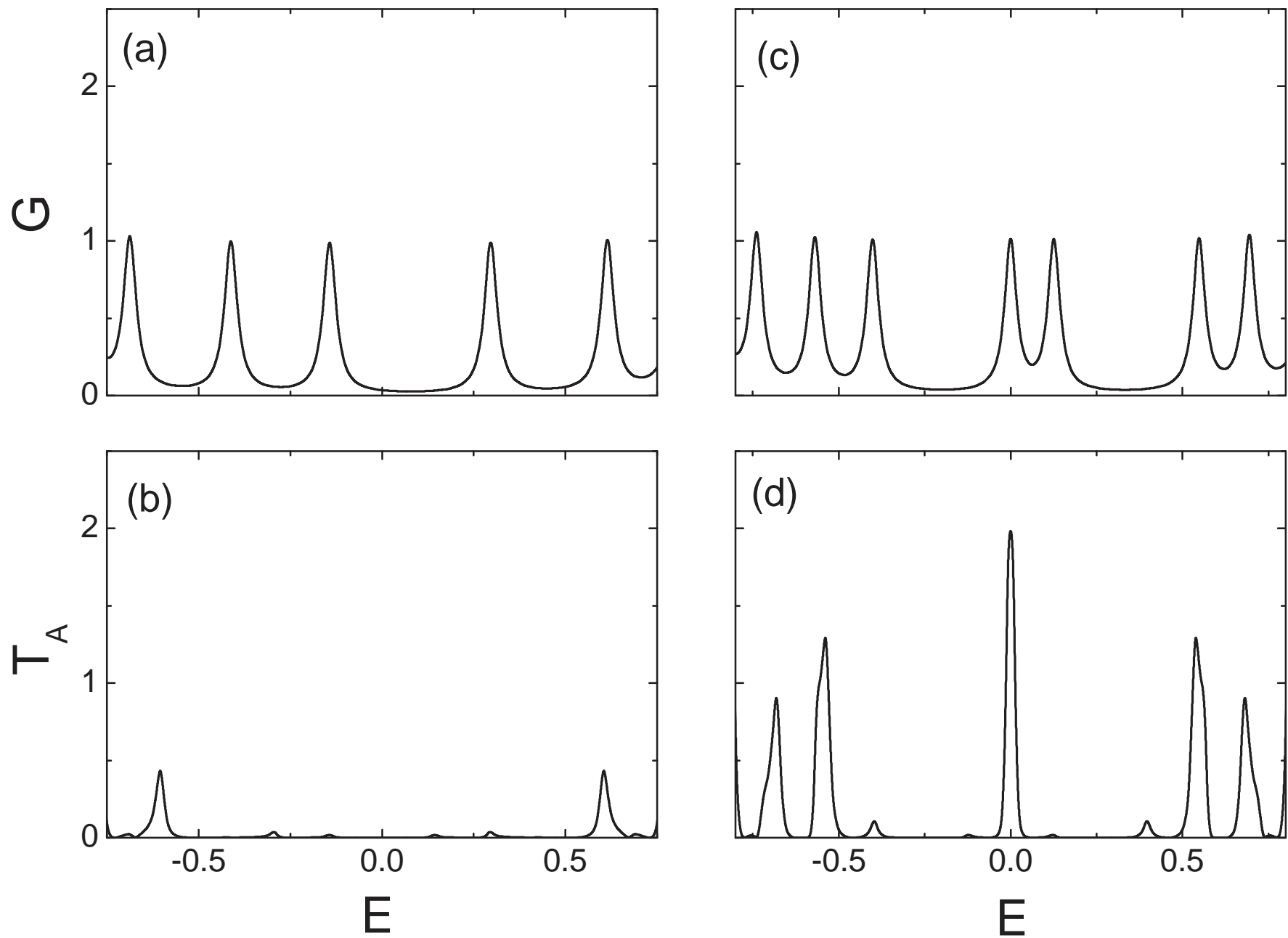
Fig. 2. (a) and (b) are the conductance $G$ and Andreev
reflection probability $T_{A}$ as a function of electron energy
$E$ for the N/CNT/N and N/CNT/S systems of $L=3$ with a boron
impurity, respectively. (c) and (d) are the corresponding $G$ and
$T_{A}$ for the systems of $L=4$, respectively.

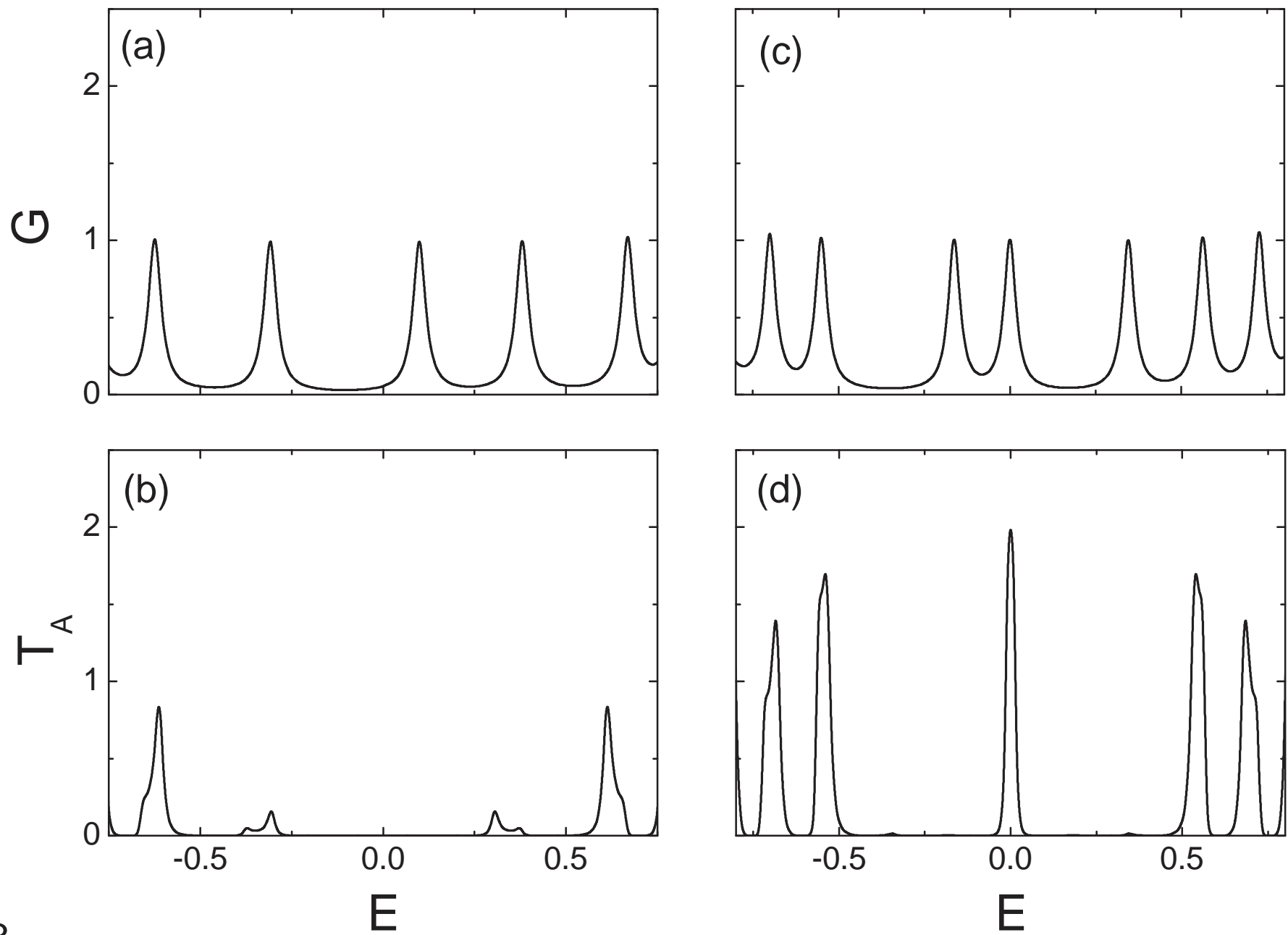
Fig. 3. (a) and (b) are the conductance $G$ and Andreev
reflection probability $T_{A}$ as a function of electron energy
$E$ for the N/CNT/N and N/CNT/S systems of $L=3$ with a nitrogen
impurity, respectively. (c) and (d) are the corresponding $G$ and
$T_{A}$ for the systems of $L=4$, respectively.

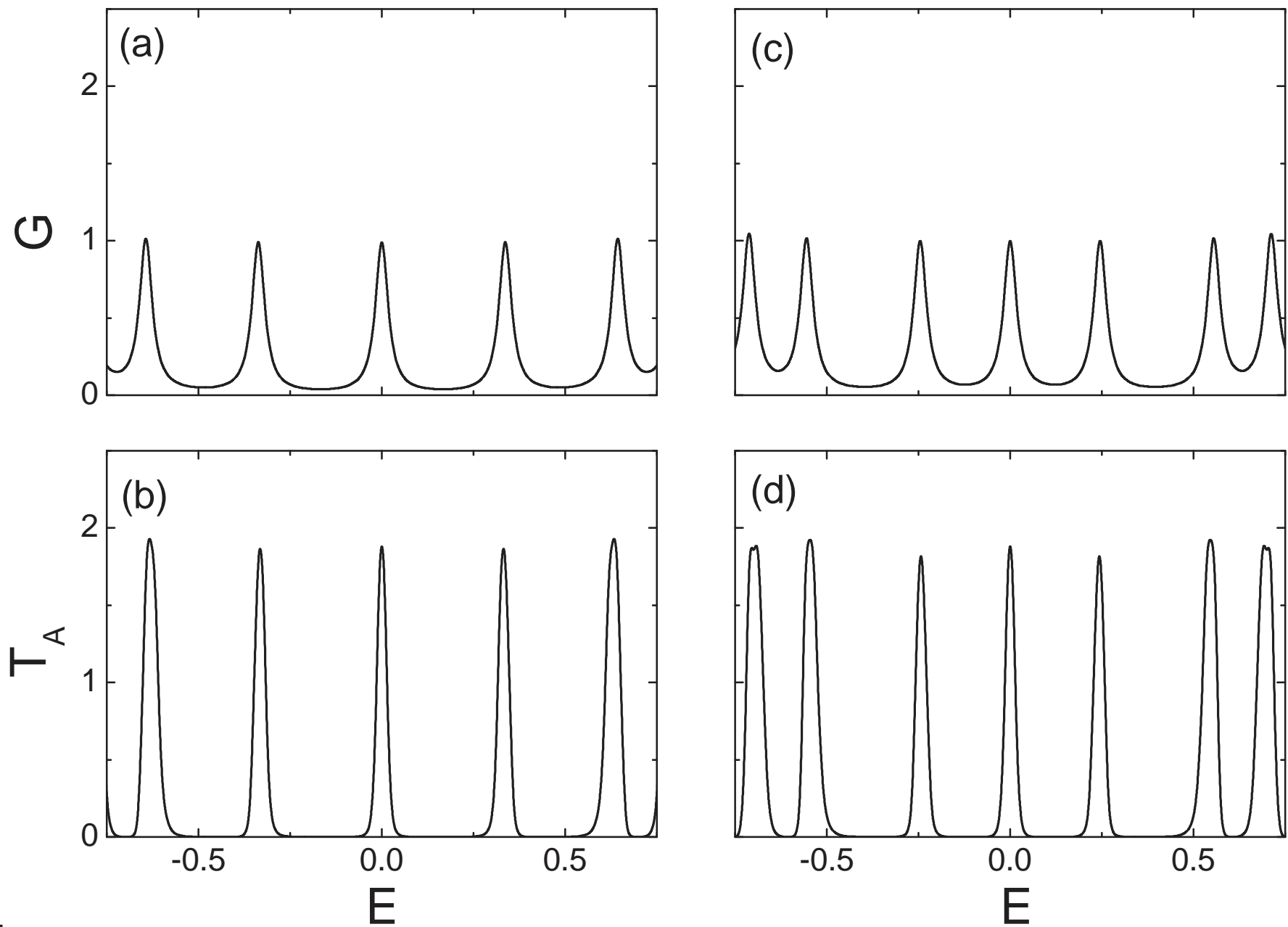
Fig. 4. (a) and (b) are the conductance $G$ and Andreev
reflection probability $T_{A}$ as a function of electron energy
$E$ for the N/CNT/N and N/CNT/S systems of $L=3$ with a vacancy,
respectively. (c) and (d) are the corresponding $G$ and $T_{A}$
for the systems of $L=4$, respectively.

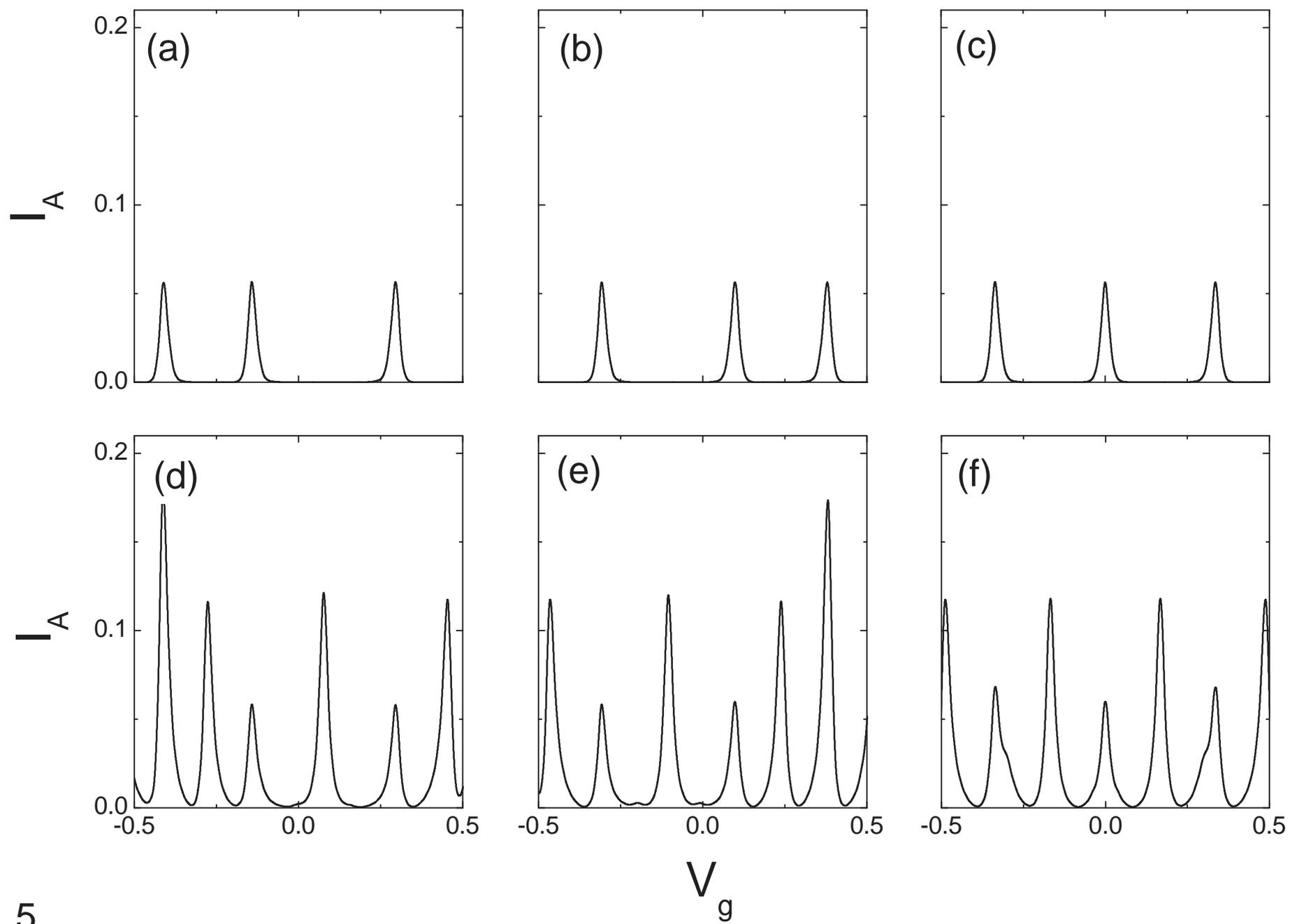
Fig. 5. (a)-(c) are the Andreev reflection current $I_{A}$ as a
function of the gate voltage $V_{g}$ at the bias $V=0.03$ with a
boron, a nitrogen, and a vacancy, respectively. (d)-(e) are the
corresponding $I_{A}$ at the bias $V=0.3$. Here $L=3$.

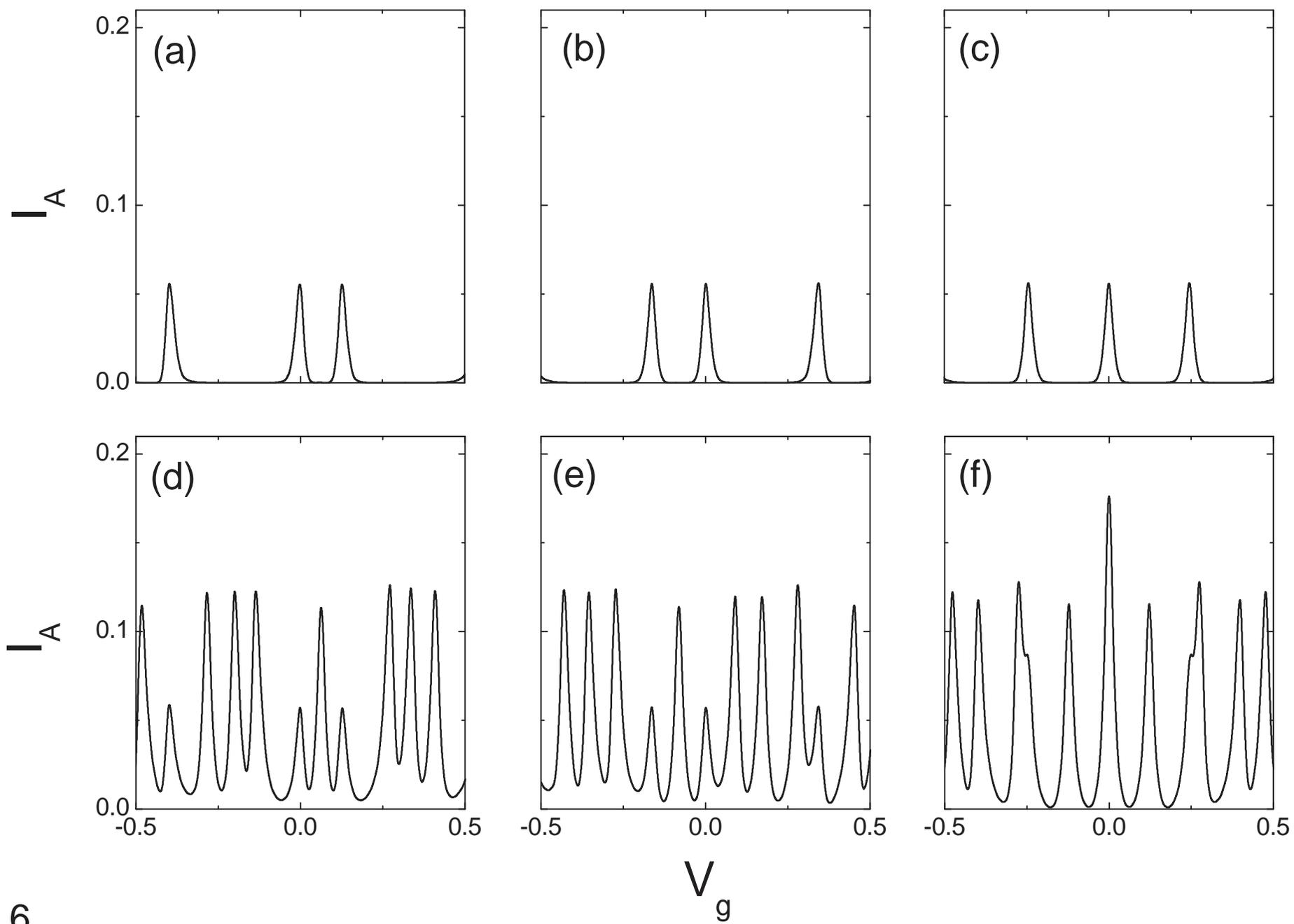
Fig. 6. (a)-(c) are the Andreev reflection current $I_{A}$ as a
function of the gate voltage $V_{g}$ at the bias $V=0.03$ with a
boron, a nitrogen, and a vacancy, respectively. (d)-(e) are the
corresponding $I_{A}$ at the bias $V=0.3$. Here $L=4$.

\end{document}